\documentclass[rnote]{aa} 

\usepackage{graphicx}
\usepackage{txfonts}
\begin{document}

   \title{X Ray detection of GJ 581 and  simultaneous UV observations}


\author{Vincenzo Vitale\inst{1}\thanks{vincenzo.vitale@roma2.infn.it}
  \and Kevin France\inst{2}}

\institute{ASI Science Data Center \& Istituto Nazionale di Fisica Nucleare,
  Via della Ricerca Scientifica, 1 00133 – Roma - Italy
  \and Center for Astrophysics and Space Astronomy, 
University of Colorado, 389 UCB, Boulder, CO 80309, USA}

   \date{}

 
  \abstract
   { The M3 dwarf GJ 581 hosts a rich system of exo-planets, some of which are potentially within or at the edge of the habitable zone (HZ). 
Nevertheless, the system habitability might be reduced  by large and sterilizing high energy emission flares, if these are frequent.}
   { The GJ 581 radiation environment was studied with simultaneous X-ray  and UV observations, which were 
performed with the XRT and UVOT instruments, respectively, on board of the SWIFT satellite.}
   { X-ray and UV data were analysed with the distributed standard tools.}
   {The dwarf GJ 581 was detected for the first time in the  0.2–10 keV range   
with an intensity of  (8$\pm$2)$\times$10$^{-4}$cts/s and a signal-to-noise ratio of 3.6. 
If black-body or APEC spectra are assumed, then the source X-ray flux is found to be 
between  1.8 and 3.3$\times$10$^{-14}$erg cm$^{-2}$s$^{-1}$ and log$_{10}$(L$_{X}$) between   25.914 and 26.176.
Despite hints of X-ray variability, better statistics are needed to establish  robust evidence for this property.
The UV measurements, obtained during 13 pointings, are also reported.
A combination of these Swift X-ray and  Hubble Space Telescope UV  measurements (with Lyman-alpha) indicate a low X-ray to UV luminosity ratio  of  $\sim$~4\%.}
   {Simultaneous X-ray and UV observations of GJ 581 are reported.
These constitute an experimental view of the system radiation environment, which will be a 
useful input for the  habitability studies of the GJ 581 planetary system.}


   \maketitle
%

\section{Introduction}

Life "as we know it" is found on a rocky planet surface, protected by an atmosphere,   within the  
planetary system HZ  (Kasting et al. 1993).
For this reason the study of exo-planetary physical conditions, such as the radiation environment, 
has large relevance in the search for extra-terrestrial life.
An important target for this search are M stars, 
because they are  the most abundant in the solar neighbourhood (Miller and Scalo 1979) and likely also in the Galaxy. 
Concerns on the M dwarf systems habitability include:
(i) tidal locking (Dole 1964), in which these  planets should be to have liquid water within the conventional HZ.
Nevertheless, this problem was subsequently reconsidered (Joshi 2003);
(ii) large flare rates, which could give rise to relatively frequent episodes of very intense planetary irradiation  with sterilizing X-ray and UV emissions, or originate long term atmosphere evaporation.
Other effects on the planetary atmospheric chemistry of a strong stellar flares, such as the ejection of energetic particles,  also have been investigated (Segura et al. 2010).

Planetary irradiation is related to the large flares occurrence and to the atmosphere's capability of shielding.
The energy distribution of coronal flares from late-type stars was investigated in  Audard et al. (2000).
According to these authors, the energy distributions are well described by 
power laws, such as
$$ \frac{dN}{dE} \ = \ k_{1} \ E^{-\alpha}   \ with \ E >  E_{min},  \alpha  >  1,$$
where dN is the number of flares with energy within the dE energy range,
E$_{min}$ is the minimum energy below which the distribution is not valid for example because of a large change in the spectral index $\alpha$,
 and k$_{1}$ is the normalization constant.
They also considered the case when E$_{min}$ is below the instrument detection threshold.
A large fraction the X-ray  emission is then  seen as a quiescent emission (90\% in their case) 
but   originates in a  superposition of many undetectable small flares,
while the rest  is seen as individual detectable flares.

 It was demonstrated  in Smith et al. 2004 that  thin atmospheres (below 100 g cm$^{-2}$) can  shield typical stellar X-rays fluxes and  thick atmospheres ($>$100 g cm$^{-2}$) can also efficiently shield    $\gamma$-rays.
A large fraction of the X-ray incident energy, up to the 10\%, is redistributed into diffuse UV
with consequences on the organic chemistry.
For comparison, Earth's atmosphere allows 2$\times$10$^{-3}$ (up to 4$\times$10$^{-2}$) of the incident high energy
 radiation flux  to reach the ground in the 200–320 nm range.

The high energy radiation environment also determines the exo-planetary photochemistry.  
It has been shown in Segura et al. (2005) that the spectral distributions of the  parent stars  in the ultraviolet
 have significant influence on the presence of 
proposed  bio-markers, such as  CH$_{4}$, N$_{2}$O, and CH$_{3}$Cl, in exo-planetary atmospheres.

In this note, we report  X-ray and UV emission measurements of  GJ 581, an M3 dwarf, which 
hosts a prominent planetary system with at least four confirmed planets  and possibly two others.
The planet GJ 581d is a super-earth with a mass of 5.6$\pm$0.6M$_{earth}$ and is close to the outer edge of the HZ,
within 0.11 and 0.21 AU 
(von Bloh et al. 2007, Selsis et al. 2007, Wordsworth et al. 2010, von Braun 2011).
The dwarf GJ 581 was not detected in X-ray  so far, at least to our knowledge.
An upper limit of 26.89 erg/s on the X-ray luminosity in the range between 0.1-2.4 keV was obtained with ROSAT observations (Poppenhaeger et al. 2010). 
UV observations were recently  performed  with the Hubble Space Telescope (France et al. 2013).

The note is organized as follows.
In Section 2, the used data and the analysis procedures are described and the X-ray detection is reported.
In Section 3, a short discussion of the reported measurements is given and is proposed an
approximate bound to the large X-ray flares occurrence, which is valid only under certain assumptions.
The X-ray to UV luminosity ratio is also derived as  the X-ray and MgII surface fluxes and their 
implications on the source age are discussed.

\section{Observations and data analysis}

The source was observed with the XRT and UVOT telescopes on board of the Swift satellite (Gehrels et al. 2004)
between December 2012 and March 2013. Twelve observations with exposures from 700s to 12ks were performed
within the Fill-In Targets program at observation cycle 8. 
Observation logs are in Table 2.

\subsection{XRT data reduction}

Data were reduced  with the HEASoft V6.12 package \footnote[1]{http://heasarc.nasa.gov/lheasoft/}
 and with the calibration files which were issued on March 2012
and January 2013 for the XRT and UVOT instruments, respectively.

\begin{figure}
   \centering
   \includegraphics[scale=0.225]{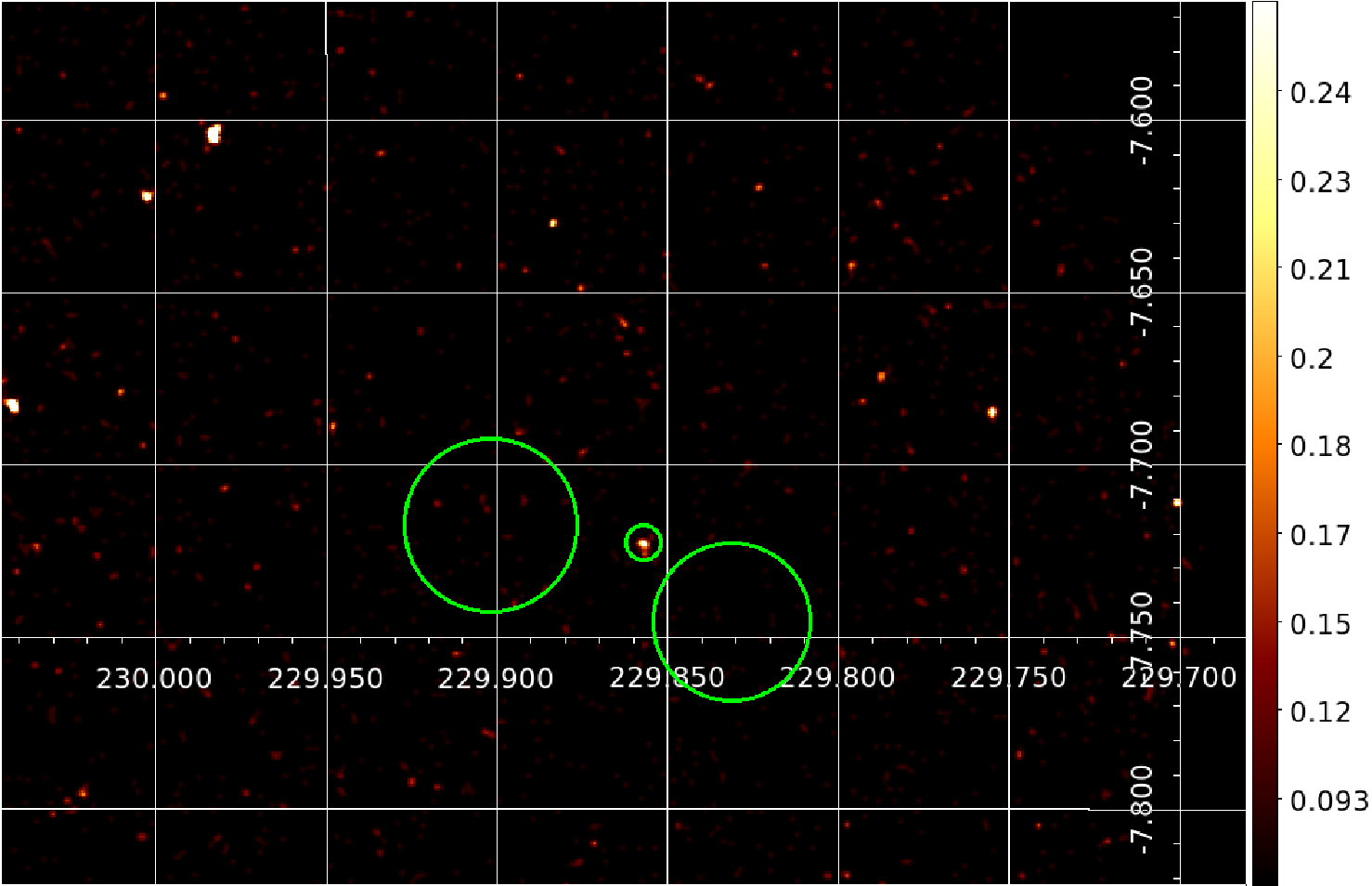}
      \caption{XRT sky map of observation 9. The smallest circle surrounds the source, which is detected with a signal-to-noise ratio
above 3. The other two larger circles are the control region areas.}
         \label{sky}
   \end{figure}

The XRT observations were carried out using the photon counting readout mode.
For XRT, the distributed level 2 cleaned event files were used with  energy between 0.2 and 10 keV and grades from  0 to 12.
Individual pointings were summed with the XSELECT tool to have a cumulative image with an exposure of 32798s.
The X-ray image was subsequently analysed with the XIMAGE tool.
An  excess was detected at RA=15 19 25.6 Dec=-07 43 21.5, which is compatible with the source location.
It consisted of 27$\pm$7 counts (which were obtained with the XIMAGE sosta tool, after correcting for various effects) 
and an intensity of (8$\pm$2)$\times$10$^{-4}$cts/s with a signal-to-noise ratio of 3.6. 
Individual pointings were grouped then into three datasets with similar exposure to investigate possible source variability.
The obtained results are listed in Table 3.
The source was detected only in the second period. The related X-ray image is shown in Fig \ref{sky}. 
During period one and three, the  count rate at the source position was  2.7  and 6.1$\times$10$^{-4}$cts/s with a low signal-to-noise ratio in both cases.
For these periods, the calculation of the count rate upper limit was then performed (XIMAGE uplim tool).
For such a calculation the used  source region was a circle with radius of 18".

The background rates were measured as a  function of the observation periods getting the number of 
events within a control region (with the XIMAGE counts tool) and dividing them by the exposure. 
The  background rates are reported in Table 3, while the control region is shown in Fig. 1. 

The average  count rate was converted into an X-ray flux  by means of the PIMMS v4.6 software.
Very simple emission models were  assumed  to convert count rates to energy flux. These models were
a black-body with temperatures of 3x10$^{6}$K and 10$^{7}$K,  as those reported in Schmitt et al. 1990 for M dwarfs
and an APEC model with an abundance parameter of 0.6. No further spectral studies were performed because of the 
limited statistics. The results are listed in Table 4.

\subsection{UVOT data reduction and UV variability study}
Sources with a signal-to-noise ratio above 3 were searched for into the UVOT images with the UVOTDETECT tool.
The dwarf GJ 581 was detected during all the  observations at the expected location with the exception of the shortest one.
Both a source region and  background  were defined on the basis of the  UVOTDETECT tool results. 
The source region consisted of  a circle with centre in RA=229.85734 and  Dec=-7.7267 (J2000.0) and with radius of 5",
while the background region was composed with two annuli; 
the first annulus with a centre in RA=229.85734 and  Dec=-7.7267 (J2000.0) and with radii of 10 and  20",
the second annulus with the same centre with  radii of 80 and  110".
For each observation the source  photometry was performed by means of the  UVOTSOURCE tool.
The tool was used with the option \emph{apercorr=CURVEOFGROWTH}  to apply an approximate aperture correction
(0.02 to 0.05 mag systematic error)
however, the source region is based on current standard  photometric aperture equal to 5".
The photometry results are reported in Column 2 and 3 of Table 5.

\begin{table}
\caption{GJ 581 parameters used for the analysis of the data. The position and proper motion parameters ($\mu \alpha$, $\mu \delta$) are taken from 
the Hipparcos catalog (GJ 581 is HIP 74995). Effective temperatures T${eff}$ are those estimated in von Paris et al. 2010. See  references therein.}
\centering
\begin{tabular}{ccc}
\hline\hline
Parameter &  Value & Ref\\

\hline
type & M3 & (Udry et al. 2007) \\
RA&  15 19 26.825 (J2000)& (Perryman et al. 1997)\\
Dec&  -07 43 20.21 (J2000)&  (Perryman et al. 1997)\\
$\mu \alpha$ & -1224.55mas/yr &  (Perryman et al. 1997)\\
$\mu \delta$ &-99.52 mas/yr & (Perryman et al. 1997) \\
distance & 6.27 pc&  (Bonfils et al. 2005)\\
T$_{eff}$ & 3190 K & (von Paris et al. 2010)\\
        & 3249 K& (von Paris et al. 2010)\\
    &       3760 K & (von Paris et al. 2010) \\
radius  & (0.29$\pm$0.010)R$_{\odot}$ & (von Braun et al. 2011)\\
\hline
\end{tabular}

\end{table}

\begin{table}
\caption{Swift Observations. The central wavelength($\r{A}$) and  FWHM($\r{A}$) of the UVOT filters are
2600 and  693 for  UVW1 , 2246 and 498 for UWM2 , 1928 and 657 for UVW2  (Poole et al. 2008 and Breeveld et al al. 2011)}
\centering
\begin{tabular}{cccc}
\hline\hline
Obs& Start Time& Filter & Exposure \\
&     &        & (UVOT,XRT)\\
&     &        &      [s] \\
\hline
1&2012-12-28 06:26:59 & W2   &  864  , 873          \\
2&2013-01-27 09:27:59 & M2   &  773 ,773              \\
3&2013-01-14 04:01:59 & W1   &  676 ,677            \\
5a,b&2013-03-10 19:17:59 & M2,W2     &    5233 ,	5192        \\

6&2013-03-12 14:38:59 & W1   &        1794 ,	1789            \\
7&2013-03-13 01:43:59 & U    &            1369	, 1367         \\
8&2013-03-14 08:11:58 & W2   &       2971 ,	2963          \\
9a,b&2013-03-15 03:28:27 &  M2,W1    &         12331 ,	12318          \\
10& 2013-03-19 03:45:59 &M2     &       586 ,	585            \\
11a,b&2013-03-22 02:02:59 &  W2,M2      &           4670 ,	4659        \\
12& 2013-03-27 06:54:33        &  M2      &        812 ,	810    \\

\hline
\end{tabular}
\label{obs}
\end{table}

\begin{table}
\caption{X-Ray results. The control region is defined in subsection 2.1.}
\centering
\begin{tabular}{cccc}
\hline\hline
Obs. & Exposure  & Intensity  & Control Region\\
       & [s] &  10$^{-4}$[cts/s]   & 10$^{-4}$[cts/s] \\
\hline
 all  & 32798  & 8 $\pm$2   & 58$\pm$4 \\
 1-8  & 13612  & $<$8      & 58$\pm$6\\
  9   & 12326  & 15$\pm$5  & 53$\pm$7\\
 10-12 & 6860  & $<$17     & 70$\pm$10 \\
\hline
\end{tabular}
\label{xres}
\end{table}

\begin{table}
\caption{Unabsorbed X-Ray Flux and Luminosity. The used N(h) in the direction of GJ 581 was measured directly by France et al. 2013 to be 2.24$\times$10$^{18}$cm$^{-2}$. The assumed APEC abundance was 0.6.}
\centering
\begin{tabular}{cccc}
\hline\hline
model & kT  & flux & luminosity\\
      &   keV    & 10$^{-14}$erg cm$^{-2}$s$^{-1}$ & 10$^{25}$erg s$^{-1}$ \\
\hline
 bb &  0.27& 1.8$\pm$0.5  & 8.2$\pm$2\\
 bb &  0.86& 3.3$\pm$0.9  & 15$\pm$4\\
APEC & 0.27& 2.0$\pm$0.6  & 9.0$\pm$3\\
APEC & 0.86& 1.8$\pm$0.5  & 8.2$\pm$2\\
\hline
\end{tabular}
\label{flu}
\end{table}

%
%

The UVOT observations were performed with four different filters (UVW2,UVM2,UVW1 and U, see Poole et al. 2008 for details);
the source count rates spanned a wide range from 0.05 to more than 80  cts/s.
The results obtained with the same filter can be directly compared.
The counts rate and associated uncertainties for each observation are reported in Table 5.
The mean count rates and rms for the UVW2,UVM2, and UVW1 are, respectively: 3.17 and 0.046cts/s; 0.08 and 0.013 cts/s; and 9.4 and 0.6 cts/s.

The parameter $k$ = R$_{obs}$/R$_{expected}$, which the ratio between the observed  and the expected count rate,
was introduced to compare the different filters observations.
The expected count rate for  each used filter  was calculated, by folding a spectral model  with the filter in-orbit effective areas:
$$R_{expected} = \sum_{i}^{m}  \ \Phi(\lambda_{i})\times A_{eff}(\lambda_{i})/\epsilon(\lambda_{i})$$
where 
A$_{eff}(\lambda_{i})$ is the effective area of the current filter as a function of the wavelength, which is  taken from the CALDB file in use,
$\Phi(\lambda_{i})$ is the energy flux,
$\epsilon(\lambda_{i})$ is the photon energy,
 and the index $i$ runs over m bins.

An arbitrary spectral model was made with
(i) the experimental measurements of the ultraviolet spectral energy distribution of GJ 581, which were obtained with the Hubble Space Telescope during July 2011 and April 2012 (France et al. 2013),
in the band between 1150 and 3140 $\r{A}$ and 
(ii) a black-body emission extrapolation in the range  between 3140 and 6000 $\r{A}$. This component was normalized  to provide a  flux of 3$\times$10$^{-15}$erg cm$^{-2}$s$^{-1}$$\r{A}$$^{-1}$ at 3100$\r{A}$ for all the used temperatures.
The count rate calculation was  limited to effective collection area values above 10$^{-2}$cm$^{2}$, given that the effective area measurement had 
associated errors of 1\% (Poole et al. 2008).

The expected count rates are  sensitive to the spectral model parameters, such as  the black-body temperature and normalization.
 In von Paris et al. 2010,,  the range between T$_{min}$=3190K and  T$_{max}$=3760K  is reported 
as descriptive for  the various effective temperatures, which are found in literature.
The difference between the expected count rates obtained with T$_{min}$=3190K and  T$_{max}$=3760K are 
2\%, 5\% ,and 12\% for the UVW1, UVM2, and UVW2 filters respectively .
The average of the two expected count rates, obtained with these two extreme temperatures
is assumed as the final expected count rate.
In general, a systematic effect on the parameter $k$ is introduced
by the choice of both effective temperature and black-body normalization
because of the different wavelength coverage of the filters.
This  effect  can be limited by minimizing the difference of the $\bar{k}$(W1), $\bar{k}$(W2) and $\bar{k}$(M2),
which are  the mean values of the $k$ parameters obtained with the three UV filters.
From Table 5 it can be seen that $\bar{k}$(W1)=0.95, $\bar{k}$(W2)=1.00 and $\bar{k}$(M2)=0.78, therefore a
systematic error of at least 11\% on the parameter $k$ should be considered.
The expected rates for UVW1,UVM2 and UVW2 are respectively: 9.85cts/s, 0.097.5cts/s and 3.15cts/s.
The mean $k$ parameter and rms are respectively 0.88 and 0.13.

%
   \begin{figure}
   \centering
   \includegraphics[width=9cm,height=7cm]{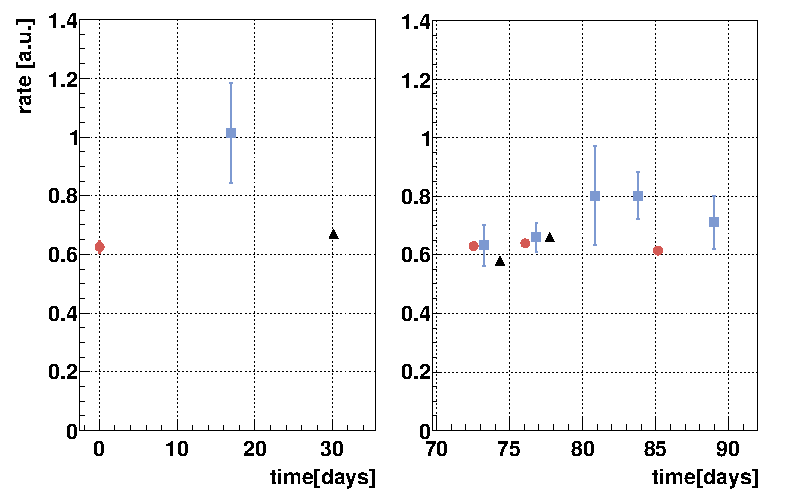}
      \caption{UVOT Lightcurve. The UVM2 rates (squares) were multiplied by 10, the UVW2 rates (circles) by 0.198, and UVW1 ones (triangles) by a factor 0.056. These two last factors are calculated by equalizing UVM2 and UVW2(UVW1) rates during observations 5 and 9, when two filters were used. For UVW1 and 2,   error bars are often smaller than the data-point marker.}
         \label{lc}
   \end{figure}
%

\begin{table}
\caption{UV measurements. Observations with the U filter are not reported.
The mean count rates and rms for the UVW2,UVM2 and UVW1 are respectively 3.167 and 0.046cts/s, 0.08 and 0.013 cts/s, 9.375 and 0.597 cts/s.
 The mean $k$ parameter and rms are respectively 0.88 and 0.13. Only statistical errors are reported for the parameter $k$ and a
systematic error of at least 11\% on this parameter  should be considered.}
\centering
\begin{tabular}{ccccc}
\hline\hline
Obs\&Filter & expo & s        & Rate &  $k$ \\
            & [s]  & [$\sigma$] & [cts/s]  &    \\
\hline
1 W2   & 850  & 48.5  & 3.16$\pm$0.09   &  1.00$\pm$0.03\\
2 M2   & 664  &  6.4  & 0.101$\pm$0.017 &  1.03$\pm$0.18 \\
3 W1   & 764  & 83.6  & 9.87$\pm$0.23   &  1.00$\pm$0.03  \\
5 W2   & 1906 & 73.8  & 3.18$\pm$0.08   &  1.01$\pm$0.03  \\
5 M2   & 3204 &  9.5  & 0.063$\pm$0.007 &  0.65$\pm$0.08 \\
6 W1   & 1761 &126.8  & 8.53$\pm$0.07   &  0.861$\pm$0.007 \\
8 W2   & 2917 & 91.9  & 3.23$\pm$0.07   &  1.02$\pm$0.02  \\
9 M2   & 7696 & 15.4  & 0.066$\pm$0.005 &  0.67$\pm$0.05 \\
9 W1   & 4427 &200.7  & 9.71$\pm$0.19   &  0.98$\pm$0.02 \\
10 M2   & 576 & 4.7   & 0.080$\pm$0.017 &  0.82$\pm$0.18  \\
11 M2   &2603 &10.5   & 0.080$\pm$0.008 &  0.82$\pm$0.08 \\
11 W2   &1983 &74.4   & 3.10$\pm$0.07   &  0.98$\pm$0.02 \\
12 M2   &1611 & 7.8   & 0.071$\pm$0.009 &  0.72$\pm$0.09 \\
\hline
\end{tabular}
\label{uvrates}
\end{table}

\section{Discussion}

X-ray emission from GJ 581 was detected for the first time.
The source detection is marginal as it is obtained with a signal consisting  of 27$\pm$7 counts
with a cumulative signal-to-noise ratio of 3.6.
If simple X-ray spectra are assumed then the source X-ray flux is found to be 
in the range between  1.8 and 3.3$\times$10$^{-14}$erg cm$^{-2}$s$^{-1}$ with an associated log$_{10}$(L$_{X}$) 
between  25.914 and 26.176.

Data were divided into three periods with exposures as balanced as possible to search for variability.
During the first period, a three standard deviation intensity upper limit of 8$\times$10$^{-4}$cts/s was obtained. 
The source was detected above a signal-to-noise ratio of 3  during the second period,
while the third period again provided  an upper limit but with weaker constraints due to the shorter exposure time.
The results of the first period deviated about two standard deviations from the measurements of period two. 
Here, a main problem is the low statistics regime. 
During period one only a handful of counts were found within the source location,
and the second period signal was at the edge of detectability.
In both cases,  small spurious effects can have large impact on the results. 
A check was done with the background rate measurements as a function of the periods. 
They are found to be steady within the errors.
Larger statistics are needed to establish  robust evidence of X-ray variability.
Furthermore, the single measurement  would not allow one to sample the flares energy distribution.

Nevertheless, approximated bounds to the star X-ray activity can be obtained if the following  hypotheses are assumed:
(i) the X-ray emission, detected during  observation window 9, is  originated by a single flare;
(ii) this flare is completely within the observation window;
(iii) the flares  have an energy distribution in the form of a power 
law with indices between 1.57 and 2.24, as  reported in Audard et al. 2000 for the studied M dwarfs,
 GJ 411, AD Leo, EV Lac, and CN Leo during 1994 and 1995.
Under these assumptions, the maximum duration of the detected flare would be  $\tau$=12ks, and  the maximum energy is E$_{f}$ = $\tau$$\times$L$_{X}$= 1.6$\times$10$^{30}$erg.
 where a benchmark value of 1.3$\times$10$^{26}$erg s$^{-1}$ for L$_{X}$ has been used.
Flares with an energy E$_{f}$ or larger would have an occurrence smaller than f$_{0}$=3.3$\times$10$^{-5}$s$^{-1}$,
or one every 30ks period.
For larger energies:  $$f(E>E') < f_{0} \times (E'/E_{f})^{-\alpha+1} $$
with $\alpha$=2.24 (or 1.57).
For the benchmark energy of E'= 10$^{32}$erg the occurrence would be lower than  1.7$\times$10$^{-7}$s$^{-1}$ (or 3.1$\times$10$^{-6}$s$^{-1}$),
respectively for the two indices, which translates to less than 6 (96) of such flares in a year.
A linear correlation between the occurrence of  10$^{32}$erg flares with the X-ray luminosity is also given in Audard et al. 2000.
This relation indicates a flare occurence between  0.8 and 1.0$\times$10$^{-7}$s$^{-1}$, with both estimates below the our proposed bounds.

It should be remarked that the proposed bounds are valid only under the assumed hypotheses and are linearly
 dependent on the considered maximum energy E$_{f}$ and the supposed flare time scale  $\tau$. 
A change of $\tau$ of a factor 10 implies a change of a factor 1/17(or  1/4 for the lower $\alpha$ value) of the occurrence bound.

Relationships between age, rotation, and coronal activity for M stars were proposed by Guinan and Engle (2009)  and  Stelzer et al. (2013). Guinan and  Engle (2009) have found  that L$_{X}$ $<$ 1.5$\times$10$^{26}$erg s$^{-1}$ are related to an M dwarf with an age larger than  5 Gyr, which are older than  Proxima Cen(M5) or IL Aqr (M4). Similarly, Engle and Guinan (2011) estimated that the age of GJ 581 is 5.7$\pm$0.8 Gyr. This results was  obtained with a rotation-age relation.  Taking the X-ray luminosity-age relation presented by   Stelzer et al. (2013)   for M0-M3 stars, the X-ray luminosity of GJ 581 implies that the age of GJ 581 is larger than $\sim$4 Gyr (the last time value reported on their Fig. 15).  In both cases, the coronal activity-age relations provide constraints that are consistent with  the previous age estimates for GJ 581  ( $>$2 Gyr; Bonfils et al. 2005.)

The UV normalized count-rates vs time are reported in Fig. 2.
In the UV range, the largest flux variations are observed with the UVM2 filter, which   
most likely traces variability in the chromosphere MgII resonance doublet, the strongest emission feature in the UVM2 bandpass.
For UVM2,  the count rate rms is of  the order of 16\% of the average count rate.
The UVW1 and UVW2 observations provide smaller variations as compared the UVM2.

France et al (2013) reported a total UV luminosity (including the FUV and NUV spectral band-passes)
 for GJ 581 of  L$_{UV}$=27$\times$10$^{26}$erg s$^{-1}$ and a MgII doublet flux of 
F$_{MgII}$= 2.13$\pm$0.13$\times$10$^{-14}$erg cm$^{-2}$s$^{-1}$ obtained from Gaussian
fits to both lines of the doublet. 
Comparing our Swift X-ray observations to the existing HST data, we find the  L$_{X}$/L$_{UV}$ ratio is
0.043$\pm$0.012. However  previous results argue for larger L$_{X}$/L$_{UV}$ ratios. France et al (2013) have found a 
log$_{10}$(L$_{UV}$/L$_{Bol}$) $\approx$ -4 and Guinan and Engle (2007) log$_{10}$(L$_{X}$/L$_{Bol}$) $\approx$-3. 
Therefore, we would have expected an L$_{X}$/L$_{UV}$ ratio of greater than unity, 
and our finding of $\sim$4\% suggests relatively weak coronal activity on GJ 581.

Using the interferometrically determined radius of GJ 581 (R = 0.29$\pm$0.010R$_{\odot}$; von Braun et al. 2011),
we find an X-ray surface flux of (2.0$\pm$0.7)$\times$10$^{4}$erg cm$^{-2}$s$^{-1}$
and a MgII surface flux of  (1.96$\pm$0.11)$\times$10$^{4}$erg cm$^{-2}$s$^{-1}$.
These surface fluxes can be
  compared with previously estimated values obtained by Rutten et al. 1991 
  (Figs. 1b and 1g). 
  The  found Mg II and X-ray surface fluxes are close to the  basal
  flux limits, indicative of a relatively old main-sequence star (e.g., Hempelmann et al. 1995; Schrijver 1995; Cuntz et al. 1999); this finding
  is generally consistent with the deduced age of GJ 581, which has been found to be
  larger than $\sim$ 4 Gyr (Selsis et al. 2007).

  Another approach pertaining to the relationship between the empirical Mg II
  surface flux and the stellar age has been given by Cardini \& Cassatella 2007.
  Following their Eq. (8) with data from their Table 2, it is found that the
  implied age of GJ 581 is very large,  possibly beyond 10 Gyr.
For further comparison we considered  the  M dwarfs studied  in  Walkowicz et al. (2008). 
Compared to the MgII/X-ray ratios of other M dwarfs (e.g., Fig. 5 from Walkowicz et al. (2008)), we find 
that GJ 581 would be comparable to GJ 876(M4.0), GJ 273(M3.5),
and GJ 191(M1.0).

GJ 876 has an estimated age between 0.1 and 5 Gyr (Correia et al. 2010), and despite  has a low-to-intermediate activity 
level based on its optical spectrum has been shown to produce significant
UV flux in its HZ. This source provides  $>$50\% of the solar luminosity
received at 1 AU in the FUV band-pass (1160 - 1790 $\r{A}$, including Lyman-alpha;  France et al. 2012).
Therefore, the L$_{X}$/L$_{UV}$ ratio is mainly indicative of the hard radiation content and does not
rule out the possibility of strong UV emission in the HZ around GJ 581.

We did not attempt to discuss the habitability of GJ 581 planetary system because 
it would need a detailed planetary thermal evolution models which follows atmospheric and lithosperic phenomena, as this is  beyond the scope of the present note.

\section{Conclusions}

We present the first X-ray detection of the exoplanet host star GJ 581.  The observations performed were part of the Fill-In Targets program during SWIFT cycle 8. These simultaneous X-ray and UV observations provide an experimental view of the energetic radiation environment of the GJ 581 planetary system, which is an important input for habitability studies of the exo-planets orbiting this low mass star. 

The low value of the found L$_{X}$ suggests that GJ 581 is older than 4 or 5 Gyr, assuming the coronal activity-age relations of Stelzer et al. (2013) or Guinan and Engle (2009).  The data suggest that the X-ray emission may be variable; however, further observations with longer exposure times are required  to establish robust evidence of X-ray variability. 

Simultaneous UV photometric observations permitted monitoring of the UV variability during the X-ray observation.  We find a low value of the X-ray to ultraviolet luminosity ratio (L$_{X}$/L$_{UV total}$) when comparing the X-ray data to the MUSCLES M dwarf UV radiation field database. We detect evidence of chromospheric activity in the UV photometry, as seen by large amplitude  variation in the SWIFT UVM2 count rates when compared to those from UVW1 and UVW2.

 X-ray and MgII surface fluxes were derived,  and they imply a stellar age ($>$ 4 Gyr), which is consistent with  the age required by coronal activity-age relations and the previous estimates in literature.

\section{Acknowledgments}

The authors wish  to thank the SWIFT team for the performed observations.
We acknowledge the use of public data from the Swift data archive.
This research has made use of the XRT Data Analysis Software (XRTDAS)
 developed under the responsibility  of the ASI Science Data Center (ASDC), Italy.
This work has made use of the MUSCLES M dwarf UV radiation field database. We are grateful to our anonymous reviewer for his/her constructive comments on our manuscript.
V.V. is  grateful to Dr. M. Capalbi and Dr V.D'Elia for help and support with the XRT data analysis,
as also with the  UVOT team and help-desk colleagues for the support with the UV observations. 
V.V. acknowledges  the support of the ASI Science Data Center (ASDC), Prot. Agg. Convenzione Quadro ASI-INFN C/011/11/1.

\end{document}